# The relativistic J-matrix theory of scattering: an analytic solution

A. D. Alhaidari
***Physics Department, King Fahd UPM, Box 5047, Dhahran 31261, Saudi Arabia***
E-mail: haidari@mailaps.org

The relativistic J-matrix is investigated in the case of Coulomb-free scattering for a general short-range spin-dependent perturbing potential and in two different $L^2$ bases. The resulting recursion relation of the reference problem, in this case, has an analytic solution. The non-relativistic limit is obtained and shown to be identical to the familiar non-relativistic J-matrix. Scattering examples are given to verify the non-relativistic limit and calculate the relativistic effects in the phase shift.

## I. INTRODUCTION

The J-matrix[1-3] is an algebraic method of quantum scattering whose structure in function space parallels that of the R-matrix method in configuration space. The perturbing short range potential, $\tilde{V}(r)$, in the R-matrix method is confined to an "*R*-box" in configuration space [i.e. $\tilde{V}(r) = 0$ for $r \geq R$]. While in the J-matrix method it is confined to an "*N*-box" in function space. That is the matrix representation $\tilde{V}_{nm} = 0$ for $n,m \geq N$. In the two methods, the unperturbed (reference) problem is solved analytically enabling scattering calculation over a continuous range of energy despite the fact that confinement in both methods produce discrete energy spectra. The basis of the function space, in the J-matrix method, is chosen such that the matrix representation of the reference Hamiltonian, $H_0$, is tridiagonal. This has no parallel in the R-matrix method. It restricts the type of $L^2$ bases and limits reference Hamiltonians to those with this type of symmetry that admits such tridiagonal representations. It turns out that this restriction implies that the reference Hamiltonian must belong to the dynamical symmetry group SO(2,1)[4,5]. Recently, a simple and robust numerical scheme was developed that lift this restriction without compromising any of the advantages offered by the method[6]. The reference wave equation in the J-matrix method gives a symmetric three-term recursion relation for the expansion coefficients of the unperturbed wave function. The regularized analytic solutions of this recursion give the asymptotic scattering states that enable calculation of the phase shift after the introduction of the perturbing potential $\tilde{V}(r)$. The method together with its multi-channel extension[7,8] has been used successfully in a large number of scattering problems in atomic and nuclear physics.

P. Horodecki introduced a relativistic extension of the theory for the Coulomb free interaction[9]. This was followed by a systematic development of the theory for the Dirac-Coulomb problem that includes perturbing short-range potentials with spin-dependent coupling[10]. In this article, we investigate the case where $Z = 0$ of the latter development and in two different $L^2$ bases: the Laguerre-type and oscillator-type functions. The motivation for this treatment comes from the fact that the resulting three-term recursion relation, in this case, has an analytic solution. This is in contrast to the general case, $Z \neq 0$, where the recursion relation was too difficult to solve analytically and a numerical solution is obtained[10]. It turns out that in the Coulomb-free case, the recursion relation is identical to the non-relativistic one, whose analytic solution is well known[3], except for a redefinition of the energy variable. The non-relativistic limit is found and shown to coincide with the familiar non-relativistic J-matrix

The paper is organized as follows: in section II, we obtain the two-component relativistic Laguerre-type $L^2$ basis, which is compatible with the requirements of the J-matrix formalism. The recursion relation is derived and its analytic solution is obtained. The non-relativistic limit is also obtained in section II. The accuracy of the non-relativistic limit is verified in the scattering examples given in section III where we also calculate the relativistic effects in the phase shift. In the second of these two examples we study the case of a short-range potential with spin dependent coupling. In the Appendix, we repeat briefly the same treatment for the oscillator-type basis which is known to have computational advantage at higher energies and used more frequently in nuclear physics.

## II. THE RELATIVISTIC J-MATRIX IN THE LAGUERRE BASIS

In this section we derive the relativistic extension of the J-matrix theory of scattering by applying its formalism to the Dirac equation. In atomic units, the free radial Dirac equation for a two-component spinor is[10,11]

$$\begin{pmatrix} 1 & \alpha\left(\frac{\kappa}{r}-\frac{d}{dr}\right) \\ \alpha\left(\frac{\kappa}{r}+\frac{d}{dr}\right) & -1 \end{pmatrix}\begin{pmatrix}\phi(r)\\ \theta(r)\end{pmatrix}=\varepsilon\begin{pmatrix}\phi(r)\\ \theta(r)\end{pmatrix} \tag{2.1}$$

where $\alpha$ is the fine structure parameter, $\varepsilon$ is the relativistic energy, and $\kappa$ is the spin-orbit coupling parameter defined by

$$\kappa = \pm (j + ½) \text{ for } l = j \pm ½ \tag{2.2}$$

where $j$ is the total angular momentum quantum number.
Equation (2.1) relates the two components of the spinor wave function as

$$\theta(r)=\frac{\alpha}{\varepsilon+1}\left(\frac{\kappa}{r}+\frac{d}{dr}\right)\phi(r) \tag{2.3}$$

Substituting this back into equation (2.1) gives the following second-order differential equation for $\phi(r)$

$$\left[-\frac{d^2}{dr^2}+\frac{\kappa(\kappa+1)}{r^2}-\frac{\varepsilon^2-1}{\alpha^2}\right]\phi(r)=0 \tag{2.4}$$

which is analogous to the radial potential-free Schrödinger equation

$$\left[-\frac{d^2}{dr^2}+\frac{l(l+1)}{r^2}-2E\right]\chi(r)=0 \tag{2.5}$$

with the substitution

$$E=(\varepsilon^2-1)/2\alpha^2\ , \quad \text{and } l=\kappa \ \text{ or } \ l=-\kappa-1 \tag{2.6}$$

For building the relativistic J-matrix formalism, we need to construct an $L^2$ discrete representation in which the reference Hamiltonian

$$H_0=\begin{pmatrix} 1 & \alpha\left(\frac{\kappa}{r}-\frac{d}{dr}\right) \\ \alpha\left(\frac{\kappa}{r}+\frac{d}{dr}\right) & -1 \end{pmatrix} \tag{2.7}$$

is tridiagonal so that the Dirac operator $J=H_0-\varepsilon$ gives a symmetric three-term recursion relation for the expansion coefficients of the spinor wave function. The solution of this recursion relation, subject to proper initial conditions, gives the two "regularized" solutions of the relativistic wave equation (2.1) that behave asymptotically as sin($kr$) and cos($kr$), where $k$ is the energy-dependent wave number. Therefore, the J-matrix formalism can be applied giving the relativistic S-matrix after the addition of a perturbing short-range model potential $\tilde{V}(r)$ [3,10]. The $L^2$ space is spanned by the two-component radial spinor wave functions $\{\psi_n(r)\}_{n=0}^{\infty}$ whose upper component is $\phi_n(r)$ and lower component $\theta_n(r)$. The conjugate space is spanned by $\{\bar{\psi}_n(r)\}_{n=0}^{\infty}$ such that

$$\langle \bar{\psi}_n | \psi_m \rangle = \int_0^\infty \bar{\phi}_n(r)\phi_m(r)dr + \int_0^\infty \bar{\theta}_n(r)\theta_m(r)dr = \delta_{nm} \quad (2.8)$$

Now, the analogy of the second order differential equation (2.4) to the Schrödinger equation (2.5) suggests that the upper component can be taken to be the non-relativistic J-matrix Laguerre basis functions[3] with $l = \kappa$. That is,

$$\phi_n(r) = a_n (\lambda r)^{\kappa+1} e^{-\lambda r/2} L_n^{2\kappa+1}(\lambda r) \quad (2.9)$$

where $\lambda$ is the basis scale parameter and $L_n^{2\kappa+1}(x)$ is the generalized Laguerre polynomial. The normalization constant $a_n$ will be determined from the normalization condition (2.8). $\phi_n(r)$ satisfies the following differential equation

$$\left[ -\frac{d^2}{dr^2} + \frac{\kappa(\kappa+1)}{r^2} - \frac{\lambda(\kappa+n+1)}{r} + \frac{\lambda^2}{4} \right] \phi_n(r) = 0 \quad (2.10)$$

The requirement that the reference Hamiltonian matrix and the basis-overlap matrix, $\langle \psi_n | \psi_m \rangle$, be at most tridiagonal is satisfied by the following expression for the lower component

$$\theta_n(r) = C\left( \frac{\kappa}{r} + \frac{d}{dr} \right) \phi_n(r) \quad (2.11)$$

where the small component strength parameter, $C$, is nonzero and independent of $n$. This expression is also motivated by the solution of the wave equation in (2.3). Using the differential and recursion properties of the Laguerre polynomials[12], we can write (2.11) explicitly as

$$\theta_n(r) = \frac{\lambda C}{2} a_n (\lambda r)^{\kappa} e^{-\lambda r/2} \left[ (2\kappa+n+1) L_n^{2\kappa}(\lambda r) + (n+1) L_{n+1}^{2\kappa}(\lambda r) \right] \quad (2.12)$$

The orthogonal conjugate representation defined in (2.8) gives:

$$\begin{aligned} \bar{\phi}_n(r) &= \frac{(1-\zeta)/4}{\kappa+n+1} \phi_n(r) + \frac{\zeta}{\lambda r} \phi_n(r) \\ \bar{\theta}_n(r) &= \frac{(1-\zeta)/\lambda^2 C^2}{\kappa+n+1} \theta_n(r) \end{aligned} \quad (2.13)$$

where, $\zeta$ is an arbitrary constant parameter. The normalization constant obtained is

$$a_n = \sqrt{\frac{\lambda \Gamma(n+1)}{\Gamma(2\kappa+n+2)}} \quad (2.14)$$

In this basis, which is defined by the set of functions in equation (2.9) and (2.12), the matrix representation of the reference Hamiltonian $H_0$ is tridiagonal and has the following elements:

$$\begin{aligned} (H_0)_{n,n} &= 2(\kappa+n+1)\left[ 1 - (\lambda C/2)^2 (1 - 2\alpha/C) \right] \\ (H_0)_{n,n+1} &= -\sqrt{(n+1)(2\kappa+n+2)} \left[ 1 + (\lambda C/2)^2 (1 - 2\alpha/C) \right] \\ (H_0)_{n,n-1} &= -\sqrt{n(2\kappa+n+1)} \left[ 1 + (\lambda C/2)^2 (1 - 2\alpha/C) \right] \end{aligned} \quad (2.15)$$

The tridiagonal basis-overlap matrix elements, $I_{nm} = \langle \psi_n | \psi_m \rangle$, are:

$$I_{n,n} = 2(\kappa+n+1)\left[1+\left(\lambda C/2\right)^2\right]$$
$$I_{n,n+1} = -\sqrt{(n+1)(2\kappa+n+2)}\left[1-\left(\lambda C/2\right)^2\right] \tag{2.16}$$
$$I_{n,n-1} = -\sqrt{n(2\kappa+n+1)}\left[1-\left(\lambda C/2\right)^2\right]$$

These will define the matrix elements of the tridiagonal Dirac operator $J_{nm}(\varepsilon) = \left(H_0\right)_{nm} - \varepsilon I_{nm}$ that gives the symmetric three-term recursion relation. The expansion coefficients of the spinor wave function that solves the reference equation (2.1) satisfy this recursion relation which reads

$$J_{n,n-1}h_{n-1} + J_{n,n}h_n + J_{n,n+1}h_{n+1} = 0 \;; \qquad n \geq 1 \tag{2.17}$$

where $h_n$ stands for either $s_n$ or $c_n$ (the asymptotic sine-like and cosine-like solutions, respectively). The initial relations are[10]

$$\begin{aligned} &J_{00}s_0 + J_{01}s_1 = 0 \\ &J_{00}c_0 + J_{01}c_1 = -\alpha^2 W\big/2s_0 \end{aligned} \tag{2.18}$$

where $W(\varepsilon)$ is the Wronskian of the regular and irregular solutions of the free Dirac problem:

$$W(\varepsilon) = W(\psi_{\text{reg}},\psi_{\text{irreg}}) = \psi_{\text{reg}}\frac{d\psi_{\text{irreg}}}{dr} - \psi_{\text{irreg}}\frac{d\psi_{\text{reg}}}{dr} = -\frac{2}{\alpha}\sqrt{\frac{\varepsilon-1}{\varepsilon+1}} \tag{2.19}$$

These coefficients also satisfy the Wronskian-like relation

$$J_{n,n-1}(c_n s_{n-1} - c_{n-1}s_n) = -\alpha^2 W/2 \;; \qquad n \geq 1 \tag{2.20}$$

Using the matrix elements given in (2.15) and (2.16) above, we can write the homogeneous recursion relation (2.17) as

$$\left\{2\left(n+\kappa+1\right)\frac{(\varepsilon-1)\left[1+\left(\frac{\lambda C}{2}\right)^2\right]-\frac{\lambda^2C^2}{2}\left(\frac{\alpha}{C}-1\right)}{(\varepsilon-1)\left[1-\left(\frac{\lambda C}{2}\right)^2\right]+\frac{\lambda^2C^2}{2}\left(\frac{\alpha}{C}-1\right)}\right\}h_n(\varepsilon) + b_{n-1}h_{n-1}(\varepsilon) + b_n h_{n+1}(\varepsilon) = 0 \tag{2.21}$$

where the recursion coefficient $b_n = \sqrt{\left(n+1\right)\left(2\kappa+n+2\right)}$.

This is to be compared with the well-known analytically solved non-relativistic J-matrix recursion relation in the Laguerre basis[3]

$$\left[2\left(n+l+1\right)\frac{E-\lambda^2/8}{E+\lambda^2/8}\right]h'_n(E) + b'_{n-1}h'_{n-1}(E) + b'_n h'_{n+1}(E) = 0 \tag{2.22}$$

where $b'_n = \sqrt{\left(n+1\right)\left(2l+n+2\right)}$ and $h'_n$ stands for either the sine-like or cosine-like solutions for the non-relativistic problem. Therefore, by comparing relation (2.21) with (2.22) we conclude that

$$h_n(\varepsilon) = h'_n\left(\frac{-1}{2C^2}\frac{\varepsilon-1}{\varepsilon-1+2(1-\alpha/C)}\right)\Bigg|_{l=\kappa} \tag{2.23}$$

and that the range of the small component strength parameter is $\alpha$ > C > 0. Using the analytic solution of the non-relativistic recursion, $h'_n(E)$, derived by Yamani and Fishman[3], we can write:

$$c_n(\varepsilon) = -\frac{2^\kappa}{\sqrt{\pi}} \frac{a_n}{\lambda} \frac{\Gamma(\kappa+1/2)}{(\sin\omega)^\kappa} {}_2F_1\left(-n-1-2\kappa, n+1; 1/2-\kappa; \sin^2\frac{\omega}{2}\right)$$
$$s_n(\varepsilon) = \frac{2^\kappa}{\lambda} a_n \Gamma(\kappa+1)(\sin\omega)^{\kappa+1} C_n^{\kappa+1}(\cos\omega) \qquad (2.24)$$

where ${}_2F_1(a,b;c;z)$ is the hypergeometric function and $C_n^\nu(x)$ is the Gegenbauer polynomial. The angle $\omega$ is defined by

$$\cos(\omega) = \frac{[k(\varepsilon)/\lambda]^2 - 1/4}{[k(\varepsilon)/\lambda]^2 + 1/4} \qquad (2.25)$$

where

$$k(\varepsilon) = \sqrt{\frac{-1}{C^2} \frac{\varepsilon - 1}{\varepsilon - 1 + 2(1-\alpha/C)}} \qquad (2.26)$$

The none-relativistic limit is achieved by letting $\alpha \to 0$ [i.e. $c$(speed of light) $\to \infty$], which gives $\varepsilon \cong 1 + \alpha^2 E$ and $\kappa = l$. Moreover, in the same limit, the small component, $\theta_n(r)$, of the wave function will be negligible compared to the larger component $\phi_n(r)$. That is, the small component strength parameter $C$ in equation (2.11) will be of the order of $\alpha$. Taking this limit in the relativistic recursion relation (2.21) gives

$$2(n+l+1)\left[\frac{\alpha^2 E - \frac{\lambda^2 C^2}{2}(\alpha/C - 1)}{\alpha^2 E + \frac{\lambda^2 C^2}{2}(\alpha/C - 1)}\right] h_n(\varepsilon) + b_{n-1} h_{n-1}(\varepsilon) + b_n h_{n+1}(\varepsilon) = 0 \quad (2.27)$$

This non-relativistic limit can be identified with the non-relativistic recursion relation (2.22) if we choose $C = +\alpha/2$. Therefore, the non-relativistic J-matrix limit for the case $Z = 0$ is achieved by making the following choice of parameters in the relativistic J-matrix formalism development above:

$$\alpha \to 0$$
$$\kappa = l \qquad (2.28)$$
$$C = \alpha/2$$

with the non-relativistic energy $E \cong (\varepsilon - 1)/\alpha^2$.

The solution (2.24) gives the J-matrix kinematical coefficients $\{R_n^\pm\}_{n=1}^\infty$ and $\{T_n\}_{n=0}^\infty$ defined by

$$T_n = \frac{c_n - is_n}{c_n + is_n} \quad ; \quad R_{n+1}^\pm = \frac{c_{n+1} \pm is_{n+1}}{c_n \pm is_n} \quad ; n \ge 0 \qquad (2.29)$$

These will be the coefficients that enter in the calculation of the $N^{\text{th}}$ order relativistic S-matrix[10]

$$S^{(N)}(\varepsilon) = T_{N-1}(\varepsilon) \frac{1 + g_{N-1,N-1}(\varepsilon) J_{N-1,N}(\varepsilon) R_N^-(\varepsilon)}{1 + g_{N-1,N-1}(\varepsilon) J_{N-1,N}(\varepsilon) R_N^+(\varepsilon)} \qquad (2.30)$$

where $g_{N-1,N-1}(\varepsilon)$ is the finite Green's function in the conjugate subspace[3,10] spanned by $\{\bar{\psi}_n(r)\}_{n=0}^{N-1}$ and carries the dynamical effects of the short range model potential $\tilde{V}(r)$:

$$g_{N-1,N-1}(\varepsilon) = \langle \bar{\psi}_{N-1} | \left( H_0 + \tilde{V} - \varepsilon \right)^{-1} | \bar{\psi}_{N-1} \rangle \tag{2.31}$$

Due to the fact that the basis of the $L^2$ space is non-orthogonal, care should be exercised in the calculation of the finite Green's function as outlined in Appendix B of reference[10].

Now the matrix elements of the perturbing short-range *scalar* potential $\tilde{V}(r)$ are calculated as follows:

$$\tilde{V}_{nm} = \alpha^2 \int_0^\infty \phi_n(r) \tilde{V}(r) \phi_m(r) dr + \alpha^2 \int_0^\infty \theta_n(r) \tilde{V}(r) \theta_m(r) dr \tag{2.32}$$

With the help of (2.9) and (2.12), this can be written as

$$\begin{aligned}
\tilde{V}_{nm} = {} & \frac{\alpha^2}{\lambda} a_n a_m \int_0^\infty e^{-y} y^{2\kappa+2} L_n^{2\kappa+1}(y) L_m^{2\kappa+1}(y) \tilde{V}(y/\lambda) dy + \frac{\lambda}{4} \alpha^2 C^2 a_n a_m \times \\
& \Bigg[ (2\kappa+n+1)(2\kappa+m+1) \int_0^\infty e^{-y} y^{2\kappa} L_n^{2\kappa}(y) L_m^{2\kappa}(y) \tilde{V}(y/\lambda) dy \\
& + (n+1)(m+1) \int_0^\infty e^{-y} y^{2\kappa} L_{n+1}^{2\kappa}(y) L_{m+1}^{2\kappa}(y) \tilde{V}(y/\lambda) dy \\
& + (2\kappa+n+1)(m+1) \int_0^\infty e^{-y} y^{2\kappa} L_n^{2\kappa}(y) L_{m+1}^{2\kappa}(y) \tilde{V}(y/\lambda) dy \\
& + (2\kappa+m+1)(n+1) \int_0^\infty e^{-y} y^{2\kappa} L_{n+1}^{2\kappa}(y) L_m^{2\kappa}(y) \tilde{V}(y/\lambda) dy \Bigg]
\end{aligned} \tag{2.33}$$

where $y = \lambda r$. To evaluate these integrals we utilize a scheme based on Gauss quadrature as outlined in Appendix A of reference[10]. The result of this computation is

$$\begin{aligned}
\tilde{V}_{nm} \cong {} & \alpha^2 \Big[ \sqrt{(2\kappa+n+2)(2\kappa+m+2)} F_{n,m}^{2\kappa+2} + \sqrt{nm} F_{n-1,m-1}^{2\kappa+2} \\
& - \sqrt{m(2\kappa+n+2)} F_{n,m-1}^{2\kappa+2} - \sqrt{n(2\kappa+m+2)} F_{n-1,m}^{2\kappa+2} \Big] + \left( \frac{\alpha\lambda C}{2} \right)^2 \times \\
& \Big[ \sqrt{(2\kappa+n+1)(2\kappa+m+1)} F_{n,m}^{2\kappa} + \sqrt{(n+1)(m+1)} F_{n+1,m+1}^{2\kappa} \\
& + \sqrt{(m+1)(2\kappa+n+1)} F_{n,m+1}^{2\kappa} + \sqrt{(n+1)(2\kappa+m+1)} F_{n+1,m}^{2\kappa} \Big]
\end{aligned} \tag{2.34}$$

where $F_{nm}^{\nu}$ is the Gauss quadrature integral approximation

$$\begin{aligned}
F_{nm}^{\nu} & \equiv \sqrt{\frac{\Gamma(n+1)\Gamma(m+1)}{\Gamma(n+\nu+1)\Gamma(m+\nu+1)}} \int_0^\infty e^{-x} x^{\nu} L_n^{\nu}(x) L_m^{\nu}(x) \tilde{V}(x/\lambda) dx \\
& \cong \sum_{k=0}^{M-1} \Lambda_{nk}^{\nu} \Lambda_{mk}^{\nu} \tilde{V}(\xi_k^{\nu}/\lambda) = F_{mn}^{\nu} \qquad , M > N
\end{aligned} \tag{2.35}$$

The eigenvalues $\left\{ \xi_n^{\nu} \right\}_{n=0}^{M-1}$ and corresponding normalized eigenvectors $\left\{ \Lambda_{mn}^{\nu} \right\}_{n,m=0}^{M-1}$ are associated with the $M \times M$ tridiagonal matrix

$$\begin{pmatrix} \hat{a}_0 & \hat{b}_0 & & & & & \\ \hat{b}_0 & \hat{a}_1 & \hat{b}_1 & & 0 & & \\ & \hat{b}_1 & \hat{a}_2 & \hat{b}_2 & & & \\ & & \hat{b}_2 & \times & \times & & \\ & & & \times & \times & \times & \\ & 0 & & & \times & \times & \times \\ & & & & & \times & \times \end{pmatrix} \tag{2.36}$$

whose recursion coefficients are parameterized by $\nu$ and defined as

$$\begin{aligned} \hat{a}_n &= 2n+\nu+1 \\ \hat{b}_n &= -\sqrt{(n+1)(n+\nu+1)} \end{aligned} \tag{2.37}$$

If the short-range potential has spin dependent coupling, then it will take the following general form

$$\tilde{V}(r) = \begin{pmatrix} \tilde{V}_+(r) & \tilde{V}_0(r) \\ \tilde{V}_0(r) & \tilde{V}_-(r) \end{pmatrix} \tag{2.38}$$

Then, its matrix elements will be

$$\begin{aligned} \tilde{V}_{nm} &= \alpha^2 \int_0^\infty \phi_n(r)\tilde{V}_+(r)\phi_m(r)dr + \alpha^2 \int_0^\infty \theta_n(r)\tilde{V}_-(r)\theta_m(r)dr \\ &+ \alpha^2 \int_0^\infty \phi_n(r)\tilde{V}_0(r)\theta_m(r)dr + \alpha^2 \int_0^\infty \theta_n(r)\tilde{V}_0(r)\phi_m(r)dr \end{aligned} \tag{2.39}$$

In the special case where $\tilde{V}(r) = \tilde{A}_0 f(r)$ and $\tilde{A}_0$ is a constant 2×2 symmetric matrix (i.e. the coupling parameters $\tilde{V}_\pm, \tilde{V}_0$ are constants), then (2.39) gives

$$\begin{aligned} \tilde{V}_{nm} \cong\; & \alpha^2 \tilde{V}_+ \Big[ \sqrt{(2\kappa+n+2)(2\kappa+m+2)}\, G^{2\kappa+2}_{n,m} + \sqrt{nm}\, G^{2\kappa+2}_{n-1,m-1} \\ & - \sqrt{m(2\gamma+n+2)}\, G^{2\kappa+2}_{n,m-1} - \sqrt{n(2\gamma+m+2)}\, G^{2\kappa+2}_{m,n-1} \Big] \\ & + \left(\frac{\alpha\lambda C}{2}\right)^2 \tilde{V}_- \Big[ \sqrt{(m+1)(2\kappa+n+1)}\, G^{2\kappa}_{n,m+1} + \sqrt{(n+1)(2\kappa+m+1)}\, G^{2\kappa}_{m,n+1} \Big] \\ & + \alpha^2 \lambda C \left( \frac{\lambda C}{4} \tilde{V}_- + \tilde{V}_0 \right) \sqrt{(2\kappa+n+1)(2\kappa+m+1)}\, G^{2\kappa}_{n,m} \\ & + \alpha^2 \lambda C \left( \frac{\lambda C}{4} \tilde{V}_- - \tilde{V}_0 \right) \sqrt{(n+1)(m+1)}\, G^{2\kappa}_{n+1,m+1} \end{aligned} \tag{2.40}$$

where $G^\nu_{nm} \equiv \sum_{k=0}^{N-1} \Lambda^\nu_{nk} \Lambda^\nu_{mk} f(\xi^\nu_k / \lambda) = G^\nu_{mn}$ (2.41)

## III. ILLUSTRATIVE EXAMPLES:

We consider two examples, the first is a scattering example in the presence of the short range scalar potential $\tilde{V}(r) = 10r^2 e^{-r}$. The aim of this exercise is to demonstrate the accuracy of the non-relativistic limit and give first order relativistic effects in the scattering phase shift. The chosen parameter values in the non-relativistic limit are:

$$\alpha = 10^{-3}, \kappa = l = 0, \lambda = 2, C = \alpha/2, N = 30 \tag{3.1}$$

Figure 1(a) shows the result as a plot of $\left|1 - S^{(N)}(\varepsilon)\right|$ vs. $E$. Figure 1(b) is a plot of $\left|1 - S_{NR}^{(N)}(E)\right|$ vs. $E$ obtained by the standard non-relativistic J-matrix calculation for the same problem with the same sub-parameters ($l$ = 0, $\lambda$ = 2 and $N$ = 30). The energy variables are related by $\varepsilon \cong 1 + \alpha^2 E$ . Figures 2 is a superposition of the same two plots, however, with the only exception that the fine structure parameter $\alpha$ is chosen to be large enough ($\alpha$ = 0.2) so that the relativistic effects become relevant. Aside from variations in the overall structure of the phase shift, it should be noted that resonance shift is prominent in the relativistic S-matrix. Sharp resonance, like the one seen on the graph at $E$ = 4.188 (a.u.), are shifted only slightly. The direction of the shift is also to be noted. It is not only in one direction.

The second scattering example is for a potential with spin-dependent coupling. The potential considered is

$$\tilde{V}(r) = \begin{pmatrix} 5 & -2 \\ -2 & 3 \end{pmatrix} e^{-\tau r} \tag{3.2}$$

where $\tau$ is a positive range parameter. Figure 3(a) and 3(b) show the results in an exact parallel to Figure 1(a) and 1(b) of the first example, respectively. The parameter values taken are:

$$\alpha = 10^{-3}, \kappa = l = 0, \tau = 0.4, \lambda = 2, C = \alpha/2, N = 30 \tag{3.3}$$

Similarly, the fine structure parameter is subsequently taken to be large enough ($\alpha$ = 0.2) to bring out the relativistic effects as shown in Figure 4.

## ACKNOWLEDGMENTS

The author is grateful to Dr. H. A. Yamani for very helpful and enlightening discussions.

## APPENDIX: THE RELATIVISTIC J-MATRIX IN THE OSCILLATOR BASIS

In this Appendix, we repeat briefly the same treatment that was carried out in section II for a different, however, J-matrix-compatible $L^2$ basis: the oscillator-type Laguerre functions.

The analogy of the second order differential equation (2.4) to the Schrödinger equation (2.5) suggests that we can as well take the upper spinor component to be the oscillator-type Laguerre functions[3] with $l = \kappa$. That is,

$$\phi_n(r) = a_n (\lambda r)^{\kappa+1} e^{-\lambda^2 r^2/2} L_n^{\kappa+1/2}(\lambda^2 r^2) \tag{A1}$$

The lower spinor component, which satisfies equation (2.11), can be written explicitly using the differential and recursion properties of the Laguerre polynomials, as

$$\theta_n(r) = \lambda C a_n (\lambda r)^{\kappa} e^{-\lambda^2 r^2/2} \left[ (n + \kappa + 1/2) L_n^{\kappa-1/2}(\lambda^2 r^2) + (n+1) L_{n+1}^{\kappa-1/2}(\lambda^2 r^2) \right] \tag{A2}$$

The orthogonal conjugate representation defined in (2.8) gives:

$$\bar{\phi}_n(r) = \zeta\phi_n(r) + \frac{(1-\zeta)/2}{2n+\kappa+3/2}(\lambda r)^2\phi_n(r)$$
$$\bar{\theta}_n(r) = \frac{(1-\zeta)/2\lambda^2C^2}{2n+\kappa+3/2}\theta_n(r) \qquad \text{(A3)}$$

where, $\zeta$ is also an arbitrary constant parameter. The normalization constant is

$$a_n = \sqrt{\frac{2\lambda\Gamma(n+1)}{\Gamma(n+\kappa+3/2)}} \qquad \text{(A4)}$$

In this basis, the tridiagonal matrix elements of the reference Hamiltonian are:

$$\left(H_0\right)_{nn} = 1 + \lambda^2C^2(-1+2\alpha/C)(2n+\kappa+3/2)$$
$$\left(H_0\right)_{n,n-1} = \lambda^2C^2(-1+2\alpha/C)\sqrt{n(n+\kappa+1/2)} \qquad \text{(A5)}$$
$$\left(H_0\right)_{n,n+1} = \lambda^2C^2(-1+2\alpha/C)\sqrt{(n+1)(n+\kappa+3/2)}$$

while the tridiagonal basis-overlap matrix elements are:

$$I_{nn} = 1 + \lambda^2C^2(2n+\kappa+3/2)$$
$$I_{n,n-1} = \lambda^2C^2\sqrt{n(n+\kappa+1/2)} \qquad \text{(A6)}$$
$$I_{n,n+1} = \lambda^2C^2\sqrt{(n+1)(n+\kappa+3/2)}$$

Therefore, the resulting homogeneous recursion relation in this basis, is

$$\left[\left(2n+\kappa+3/2\right) + \frac{1}{\lambda^2C^2}\frac{\varepsilon-1}{\varepsilon+1-2\alpha/C}\right]h_n(\varepsilon) + b_{n-1}h_{n-1}(\varepsilon) + b_nh_{n+1}(\varepsilon) = 0 \qquad \text{(A7)}$$

where the recursion coefficient $b_n = \sqrt{\left(n+1\right)\left(n+\kappa+3/2\right)}$. Comparing this with the non-relativistic J-matrix recursion relation in the oscillator basis[3]

$$\left[\left(2n+l+3/2\right) - \frac{2}{\lambda^2}E\right]h'_n(E) + b'_{n-1}h'_{n-1}(E) + b'_nh'_{n+1}(E) = 0 \qquad \text{(A8)}$$

where $b'_n = \sqrt{\left(n+1\right)\left(n+l+3/2\right)}$, gives the following solution of (A7) in terms of the well known analytic solution of (A8):

$$h_n(\varepsilon) = h'_n\left(\frac{-1}{2C^2}\frac{\varepsilon-1}{\varepsilon-1+2(1-\alpha/C)}\right)\Bigg|_{l=\kappa} \qquad \text{(A9)}$$

Using the analytic solution of the non-relativistic recursion, $h'_n(E)$, found by Yamani and Fishman[3], we can write:

$$c_n(\varepsilon) = (-1)^n\frac{\Gamma(\kappa+1/2)}{\sqrt{2\pi}}\frac{a_n}{\lambda}\eta^{-\kappa}e^{-\eta^2/2}\,{}_1F_1\left(-n-1/2-\kappa;1/2-\kappa;\eta^2\right)$$
$$s_n(\varepsilon) = \frac{(-1)^n}{\lambda}\sqrt{\frac{\pi}{2}}a_n\eta^{\kappa+1}e^{-\eta^2/2}L_n^{\kappa+1/2}\left(\eta^2\right) \qquad \text{(A10)}$$

where ${}_1F_1(a;c;z)$ is the confluent hypergeometric function, $\eta(\varepsilon) \equiv k(\varepsilon)/\lambda$ and $k(\varepsilon)$ is as defined in equation (2.26).

The non-relativistic limit is also obtained by taking

$$
\begin{aligned}
&\alpha \to 0 \\
&\kappa = l \\
&C = \alpha / 2 \\
&\varepsilon \cong 1 + \alpha^2 E
\end{aligned}
\qquad \text{(A11)}
$$

The matrix elements of the *scalar* short range potential needed in the calculation of the finite Green's function of equation (2.31) are

$$
\begin{aligned}
\tilde{V}_{nm} \cong \alpha^2 F^{\kappa+1/2}_{n,m} &+ \alpha^2\lambda^2 C^2 \Big[ \sqrt{(n+\kappa+1/2)(m+\kappa+1/2)} F^{\kappa-1/2}_{n,m} \\
&+ \sqrt{(n+1)(m+1)} F^{\kappa-1/2}_{n+1,m+1} + \sqrt{(m+1)(n+\kappa+1/2)} F^{\kappa-1/2}_{n,m+1} \\
&+ \sqrt{(n+1)(m+\kappa+1/2)} F^{\kappa-1/2}_{m,n+1} \Big]
\end{aligned}
\qquad \text{(A12)}
$$

where, now

$$
F^{\nu}_{nm} \cong \sum_{k=0}^{M-1} \Lambda^{\nu}_{nk} \Lambda^{\nu}_{mk} \tilde{V}(\sqrt{\xi^{\nu}_k} / \lambda) = F^{\nu}_{mn} \qquad , M > N \qquad \text{(A13)}
$$

The eigenvalues $\left\{\xi^{\nu}_n\right\}_{n=0}^{M-1}$ and corresponding normalized eigenvectors $\left\{\Lambda^{\nu}_{mn}\right\}_{n,m=0}^{M-1}$ are associated with the same tridiagonal matrix (2.36) but for a different value of the parameter $\nu$. For the *spinor* short range perturbing potential of equation (2.38) and in the case where the coupling parameters $\tilde{V}_{\pm}, \tilde{V}_0$ are constants, we obtain

$$
\begin{aligned}
\tilde{V}_{nm} &\cong \alpha^2 \tilde{V}_+ G^{\kappa+1/2}_{n,m} \\
&+ \alpha^2\lambda^2 C^2 \tilde{V}_- \Big[ \sqrt{(n+\kappa+1/2)(m+\kappa+1/2)} G^{\kappa-1/2}_{n,m} + \sqrt{(n+1)(m+1)} G^{\kappa-1/2}_{n+1,m+1} \\
&+ \sqrt{(m+1)(n+\kappa+1/2)} G^{\kappa-1/2}_{n,m+1} + \sqrt{(n+1)(m+\kappa+1/2)} G^{\kappa-1/2}_{m,n+1} \Big] \\
&+ 2\alpha^2 \lambda C \tilde{V}_0 \Big[ \sqrt{(n+\kappa+1/2)(m+\kappa+1/2)} Q^{\kappa-1/2}_{n,m} - \sqrt{(n+1)(m+1)} Q^{\kappa-1/2}_{n+1,m+1} \Big]
\end{aligned}
\qquad \text{(A14)}
$$

where

$$
\begin{aligned}
G^{\nu}_{nm} &\equiv \sum_{k=0}^{N-1} \Lambda^{\nu}_{nk} \Lambda^{\nu}_{mk} f(\sqrt{\xi^{\nu}_k} / \lambda) = G^{\nu}_{mn} \\
Q^{\nu}_{nm} &\equiv \sum_{k=0}^{N-1} \frac{\Lambda^{\nu}_{nk} \Lambda^{\nu}_{mk}}{\sqrt{\xi^{\nu}_k}} f(\sqrt{\xi^{\nu}_k} / \lambda) = Q^{\nu}_{mn}
\end{aligned}
\qquad \text{(A15)}
$$

As a third example we consider the relativistic scattering in the oscillator basis for a short-range potential barrier with spin-dependent coupling. The potential considered is

$$
\tilde{V}(r) = \begin{pmatrix} 5 & -2 \\ -2 & 3 \end{pmatrix} f(r) \qquad \text{(A16)}
$$

where

$$
f(r) = \begin{cases} \tau \;, & r \le 2\lambda^{-1} \\ 0 \;, & \text{otherwise} \end{cases} \qquad \text{(A17)}
$$

and $\tau$ is the barrier height parameter. Figure 5(a) and 5(b) show the results in an exact parallel to Figure 1(a) and 1(b) of the first example, respectively. The parameter values taken are:

$$
\alpha = 10^{-4}, \kappa = l = 0, \tau = 2.0, \lambda = 1, C = \alpha/2, N = 30 \qquad \text{(A18)}
$$

Similarly, the fine structure parameter is subsequently taken to be large enough ($\alpha$ = 0.1) producing the relativistic effects shown in Figure 6.

**FIGURE CAPTIONS:**

**FIG. 1(a).** A plot of the non-relativistic limit of $\left|1-S^{(N)}(\varepsilon)\right|$ vs. $E$ for the first scattering example in the Laguerre basis with the scalar short range potential $\tilde{V}(r)=10r^2e^{-r}$. The energy variables are related by $\varepsilon \cong 1+\alpha^2 E$ and we took the following parameter values:

$$\alpha=10^{-3}, \kappa=l=0, \lambda=2, C=\alpha/2, N=30$$

**FIG. 1(b).** A plot of $\left|1-S_{NR}^{(N)}(E)\right|$ vs. $E$ obtained by the standard non-relativistic J-matrix calculation in the Laguerre basis for the first example with the same sub-parameters:

$l$ = 0, $\lambda$ = 2 and $N$ = 30

**FIG. 2.** Superposition of the two plots in Figures 1(a) and 1(b), however, with the exception that the fine structure parameter $\alpha$ is chosen to be large enough ($\alpha$ = 0.2) so that relativistic effects become relevant.

**FIG. 3(a).** A plot of the non-relativistic limit of $\left|1-S^{(N)}(\varepsilon)\right|$ vs. $E$ for the second scattering example in the Laguerre basis with spin dependent short range potential given by equation (3.2). We took the following parameter values:

$$\alpha=10^{-3}, \kappa=l=0, \tau=0.4, \lambda=2, C=\alpha/2, N=30$$

**FIG. 3(b).** A plot of $\left|1-S_{NR}^{(N)}(E)\right|$ vs. $E$ obtained by the standard non-relativistic J-matrix calculation in the Laguerre basis for the second example with the same sub-parameters:

$l$ = 0, $\tau$ = 0.4, $\lambda$ = 2 and $N$ = 30

**FIG. 4.** Superposition of the two plots in Figures 3(a) and 3(b) where the fine structure parameter $\alpha$ is taken large enough ($\alpha$ = 0.2) to magnify the relativistic effects.

**FIG. 5(a).** A plot of the non-relativistic limit of $\left|1-S^{(N)}(\varepsilon)\right|$ vs. $E$ for the third scattering example in the Oscillator basis with spin-dependent short range potential barrier given by equations (A16) and (A17). We took the following parameter values:

$$\alpha=10^{-4}, \kappa=l=0, \tau=2.0, \lambda=1, C=\alpha/2, N=30$$

**FIG. 5(b).** A plot of $\left|1-S_{NR}^{(N)}(E)\right|$ vs. $E$ obtained by the standard non-relativistic J-matrix calculation in the Oscillator basis for the third example with the same sub-parameters:

$l$ = 0, $\tau$ = 2.0, $\lambda$ = 1 and $N$ = 30

**FIG. 6.** Superposition of the two plots in Figures 5(a) and 5(b), but with a relatively large value of the fine structure parameter ($\alpha$ = 0.1).

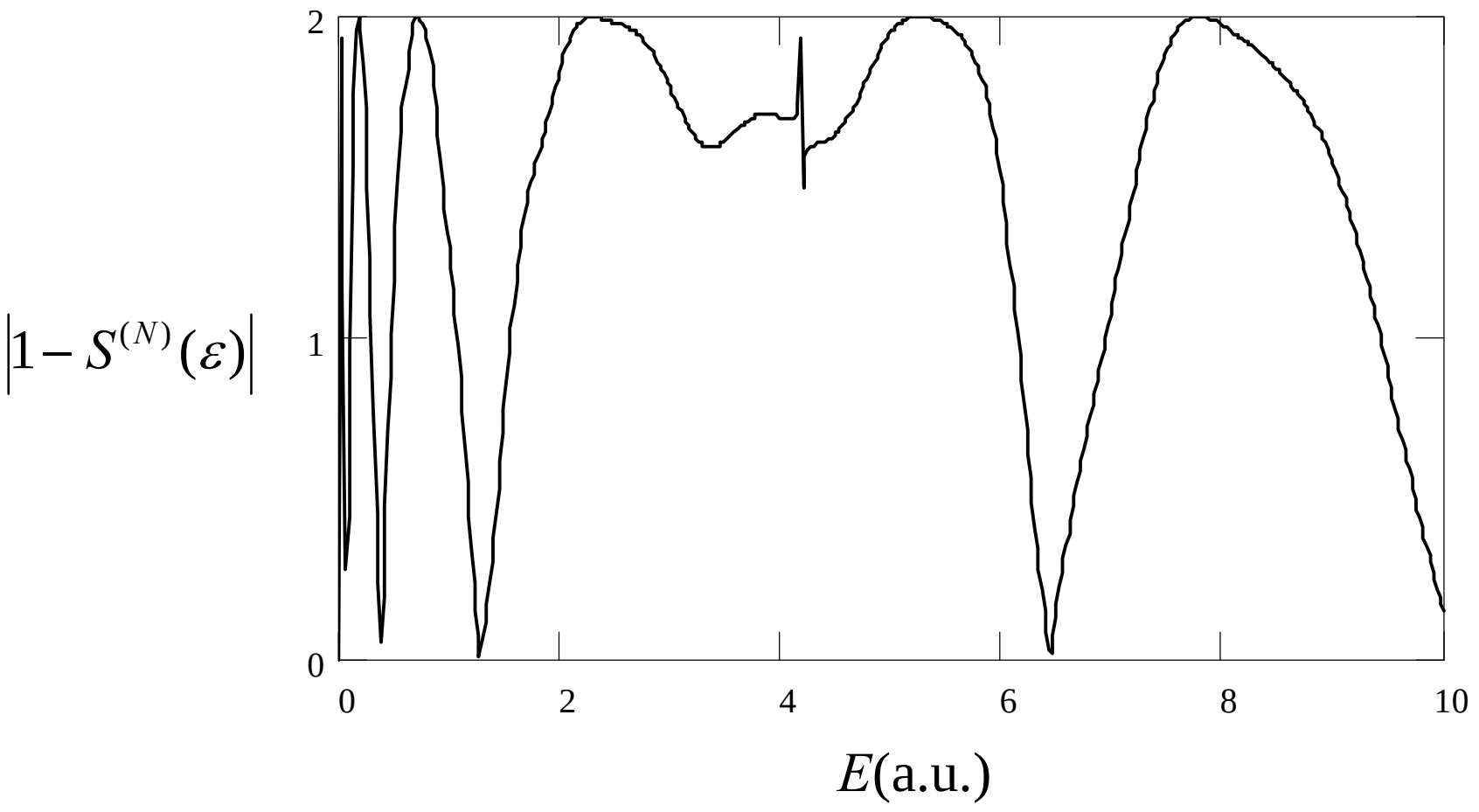

Figure 1(a)

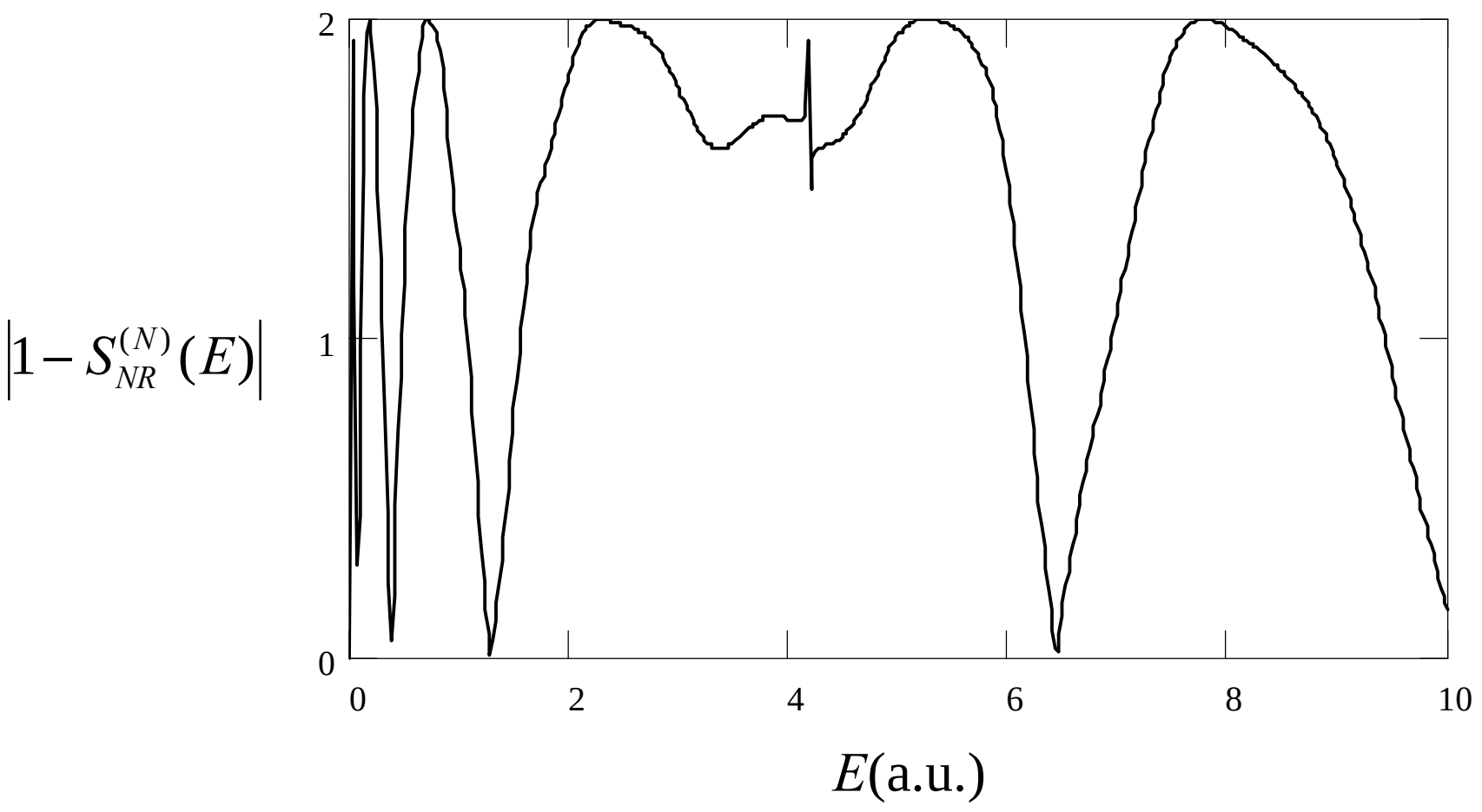

Figure 1(b)

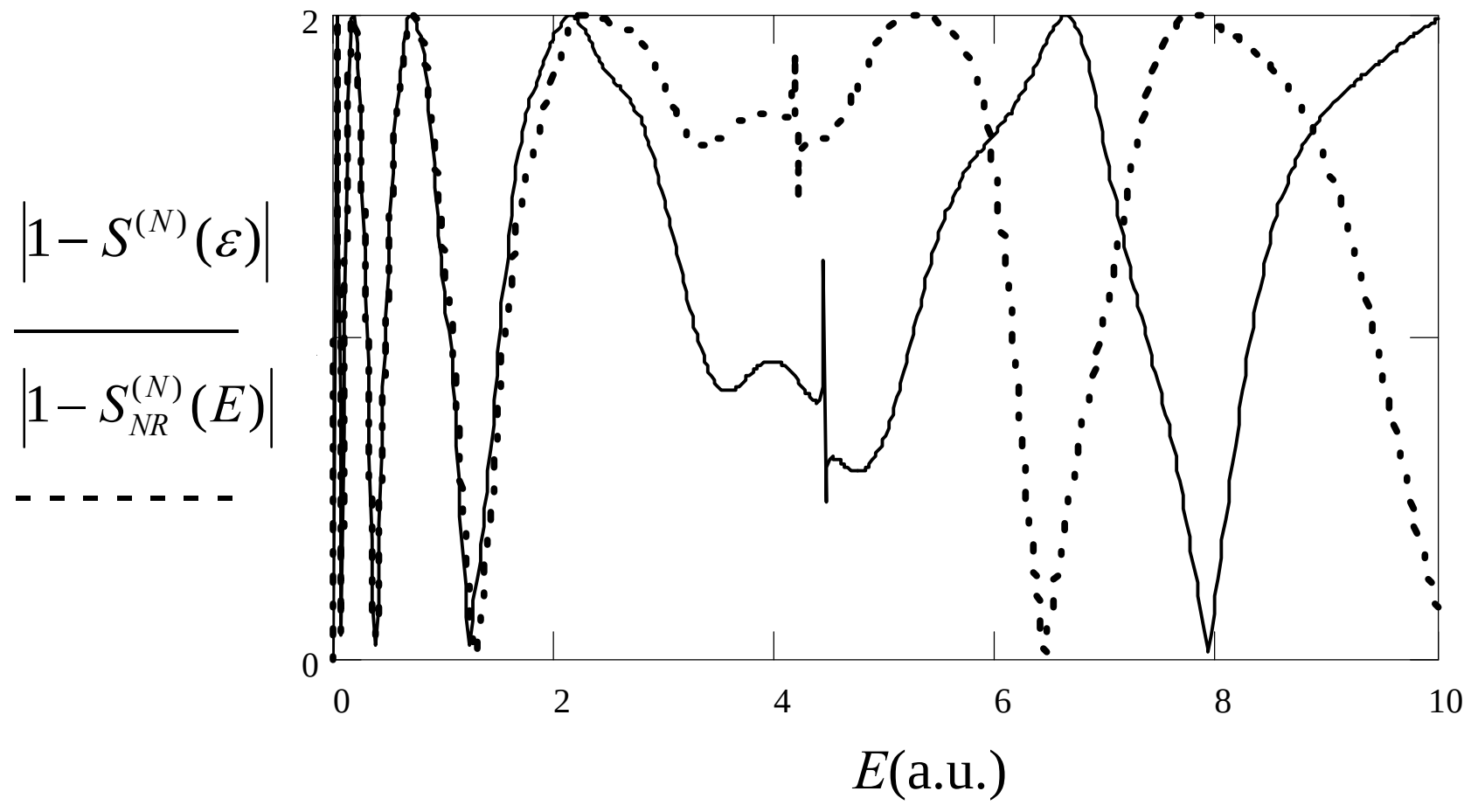

Figure 2

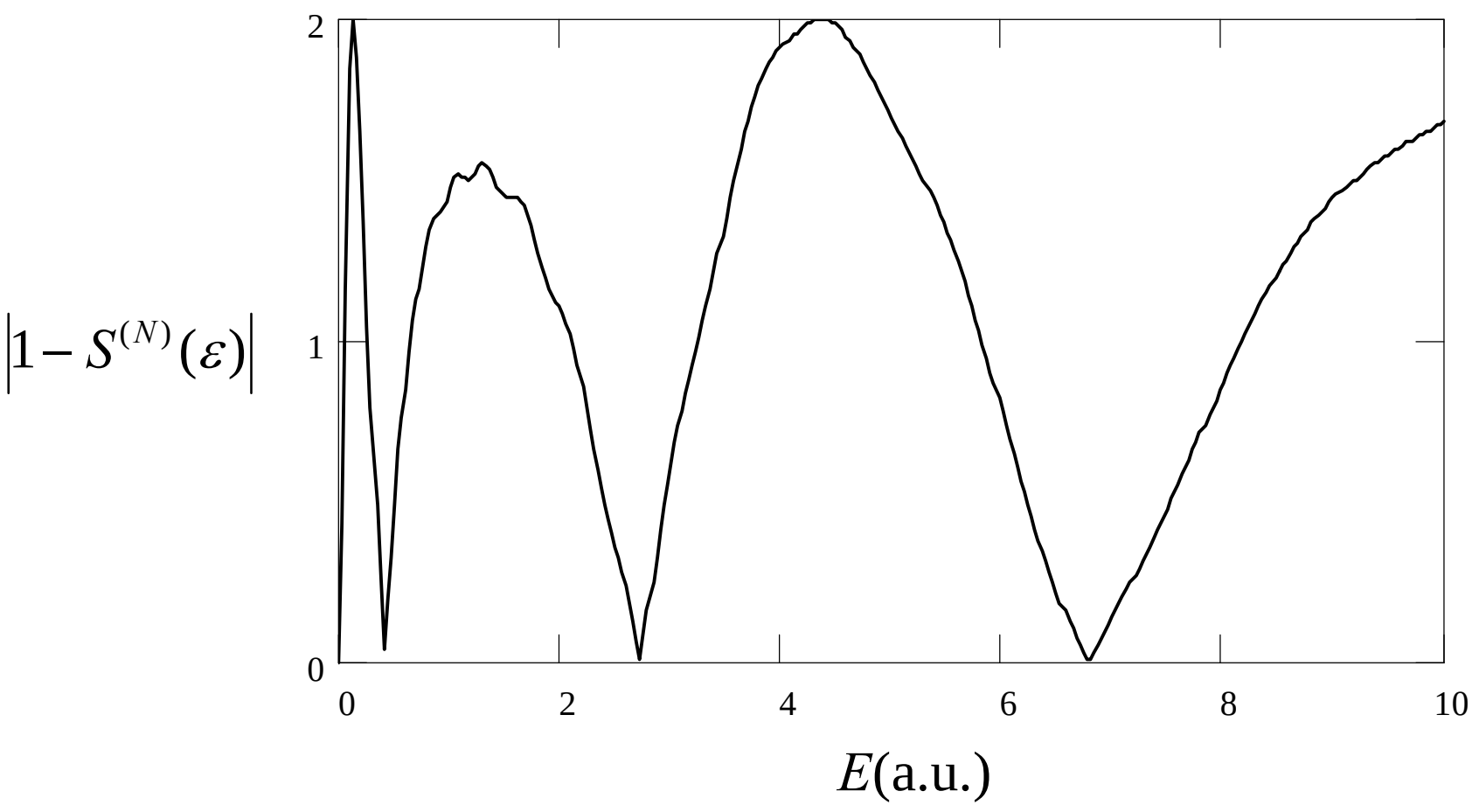

Figure 3(a)

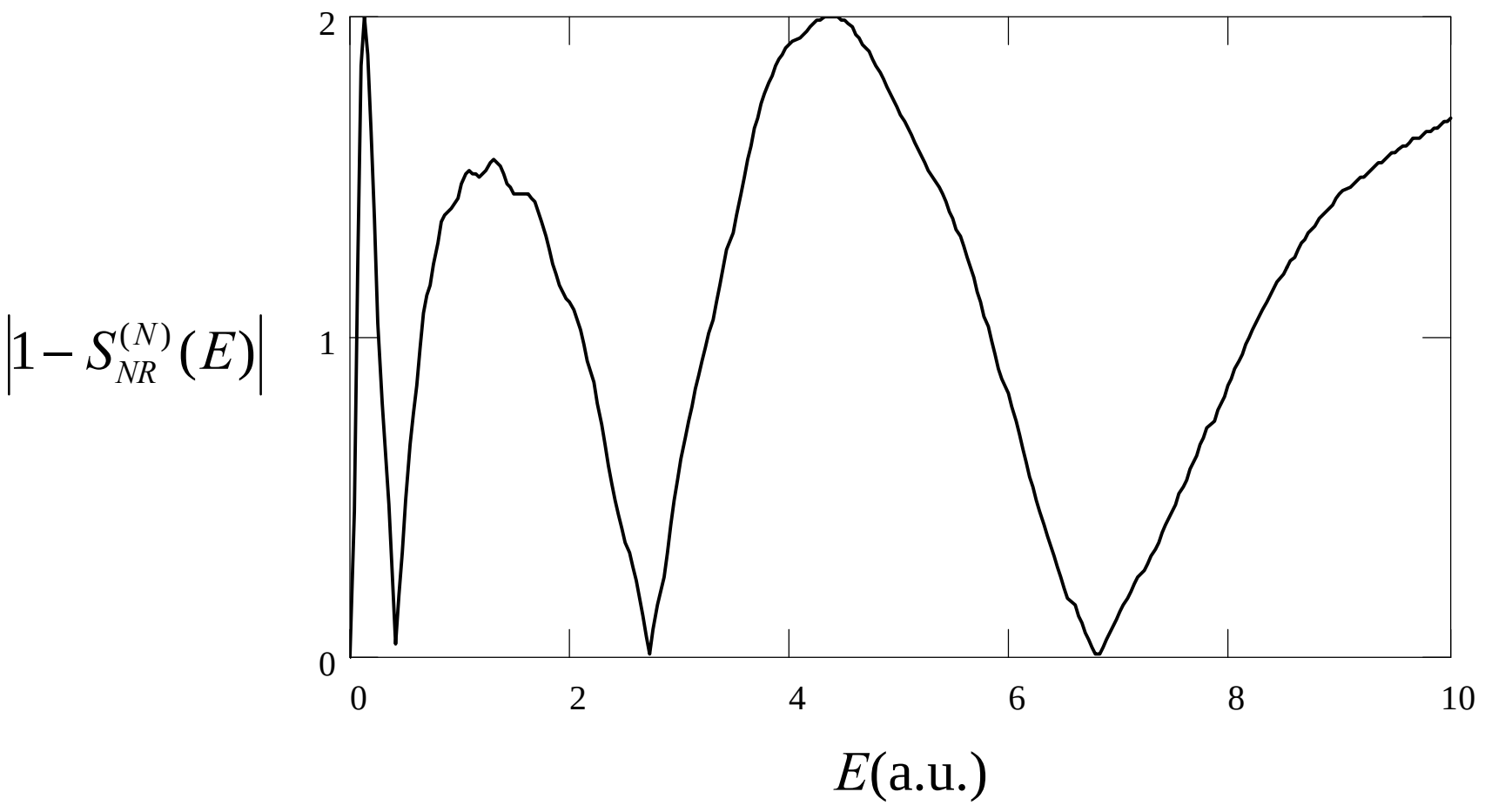

Figure 3(b)

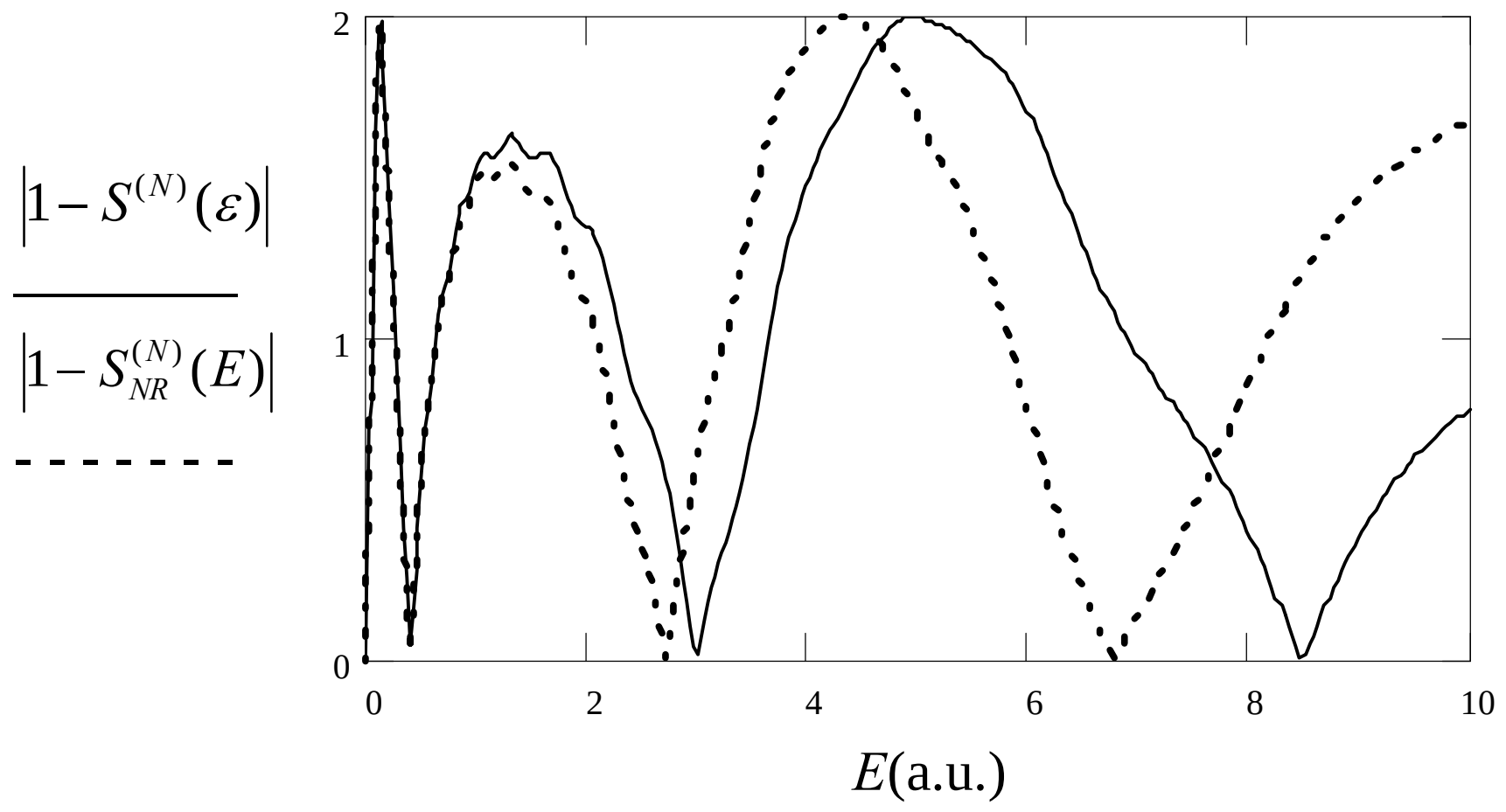

Figure 4

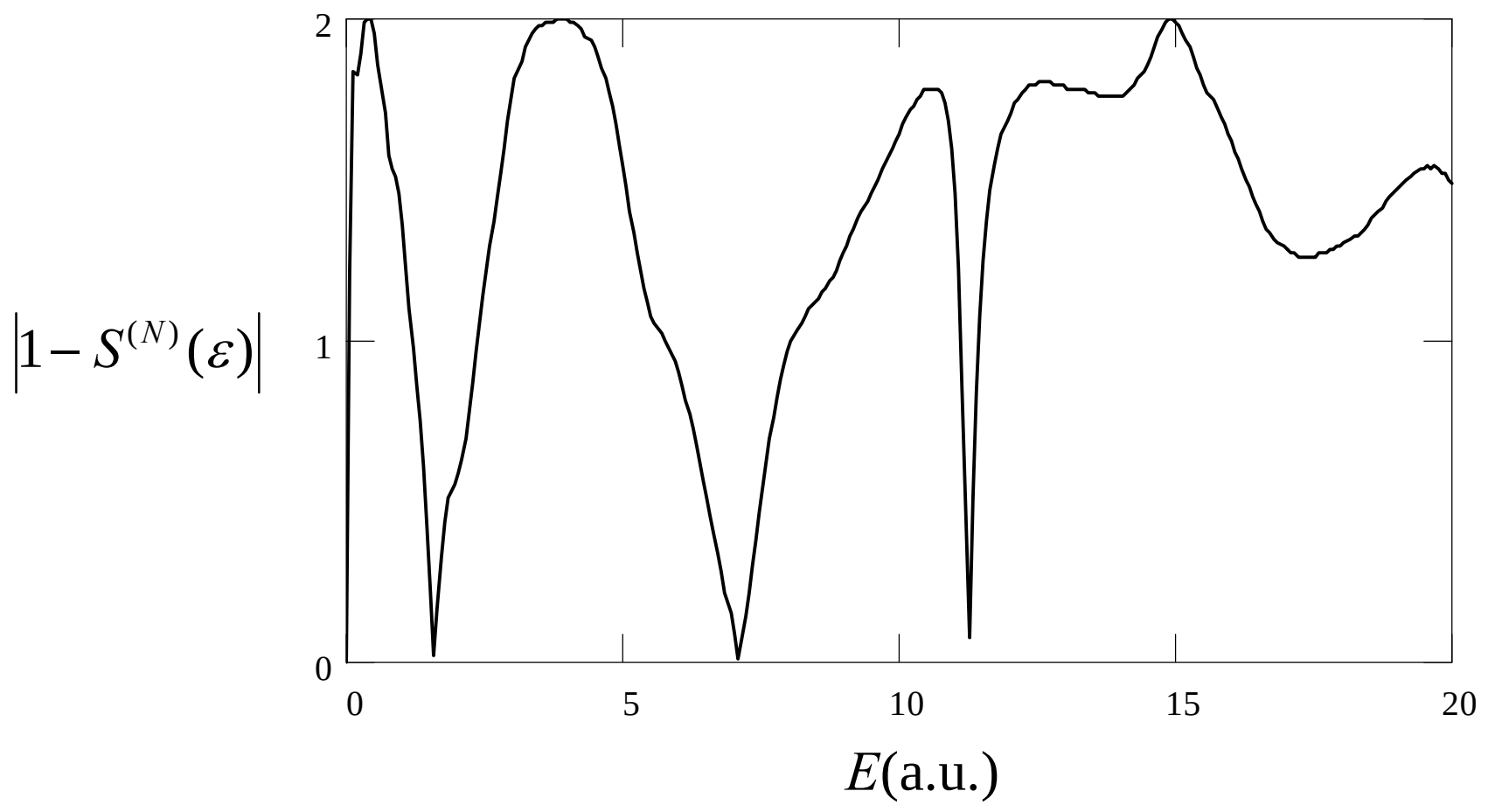

Figure 5(a)

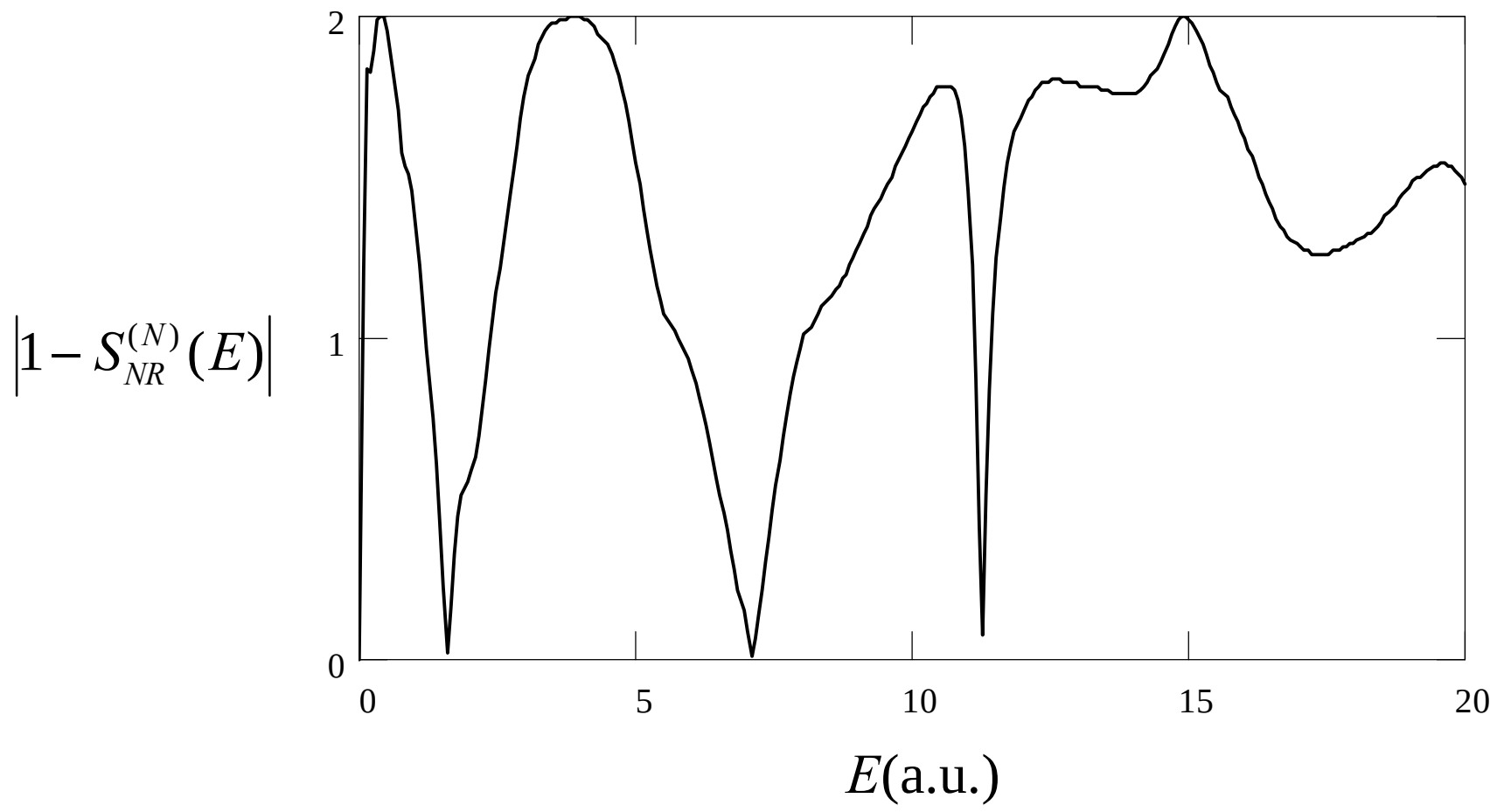

Figure 5(b)

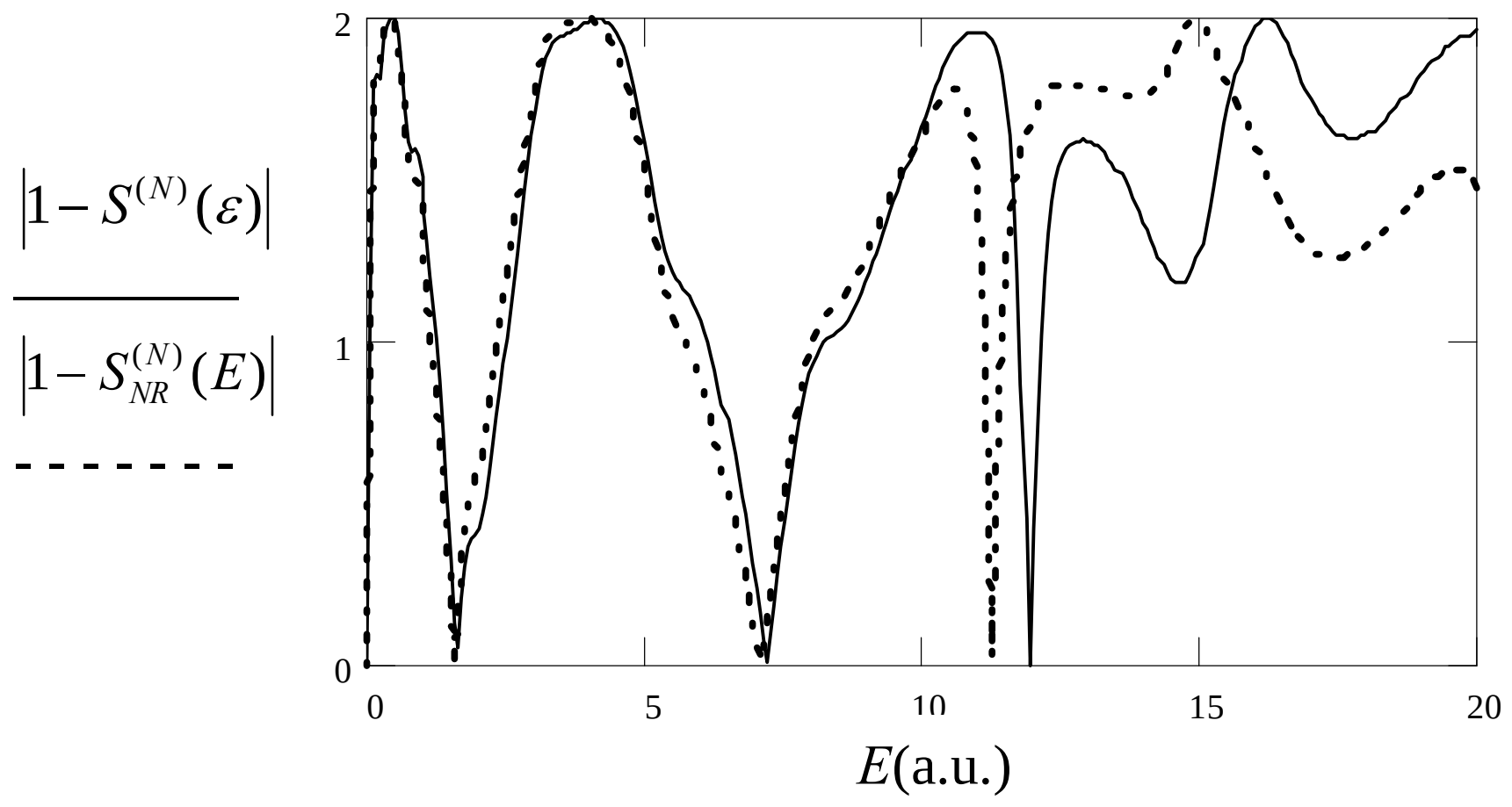

Figure 6